\documentclass[review]{elsarticle}
\usepackage{epsfig,amssymb,amsmath,float,subfig}

\newcommand{\sta}{{\rm sta}}

\newcommand{\staxn}{{\sta}_{\hspace{-.58cm}{}_{{}_{{}_{\bx\in\real^n}}}}}

\newcommand{\vsig}{\varsigma}

\newcommand{\real}{{I\!\!R}} %\newcommand{`\real}{{\bf R}}
\newcommand{\half}{\frac{1}{2}}

\newcommand{\la}{\langle}
\newcommand{\ra}{\rangle}

\newcommand{\eba}{\begin{array}}
\newcommand{\eea}{\end{array}}
\newcommand{\ebe}{\begin{eqnarray}}%[section]
\newcommand{\eee}{\end{eqnarray}}%[section]
\newcommand{\eb}{\begin{equation}}%[section]
\newcommand{\ee}{\end{equation}}%[section]

\newcommand{\calP}{{\cal{P}}}

\newcommand{\bx}{{\bf x}}

\newcommand{\calX}{{\cal X}}

\newcommand{\alp}{{\alpha}}

\newcommand{ \Lam}{{\Lambda}}

\newcommand{ \lam}{{\lambda}}

\newtheorem{thm}{Theorem}

\renewcommand\bx{{{x}}}

\newcommand\x{{{x}}}

    \newcommand\WW{W}

\renewcommand\eb{\begin{equation}}
\renewcommand\ee{\end{equation}}

\begin{document}

\title{Complete Solutions and Triality Theory to a Nonconvex Optimization Problem with Double-Well Potential
in $\real^n$}% \vspace{0.4cm}\\[0pt]

%\author{Daniel Mauricio Morales Silva \and
%        David Yang Gao} %etc.
%
%\institute{D.M. Morales Silva \and D.Y. Gao \at
% GSITMS, University of Ballarat, Ballarat, VIC 3353, Australia.
%\\\email{dmoralessilva@ballarat.edu.au} \\\email{dgao@ballarat.edu.au}
% }
%\date{Received: date / Revised: date}

%\author[rvt]{MORALES SILVA, D. M.}%\corref{cor1}}
\author{Daniel M. Morales Silva  }%\corref{cor1}}
\ead{d.moralessilva@ballarat.edu.au}
\author{David Y. Gao \corref{cor2}}%\fnref{fn1}}
\ead{d.gao@ballarat.edu.au}

%\cortext[cor1]{Principal Corresponding author}
\cortext[cor2]{Corresponding author}
%\fntext[fn1]{Fax: +61 3 5327 9077}

\address{School of Science, Information Technology and Engineering, \\
University of Ballarat,   Mt Helen, VIC 3350, Australia}

%\maketitle

\begin{abstract}
The main purpose of this research note
is to show that  the triality theory can always be used to identify
 both global minimizer and the biggest local maximizer in
  global optimization.
 An open problem left on the double-min duality  is  solved for
 a nonconvex optimization problem with double-well potential in $\real^n$, which leads to
  a complete set of analytical solutions.
  Also a convergency theorem is proved for linear perturbation canonical dual method, which
  can be used for solving global optimization problems with multiple  solutions.
  The methods and results presented in this note pave the way towards the
  proof of  the triality theory in general cases.
\end{abstract}

\maketitle

\section{Primal Problem and Motivation}
We are interested in analytical solutions to the following
global minimization problem ($(\calP)$ in short):
\eb
(\calP): \;\;\;  \min \{ \Pi(x) = \WW(x) - \la x, f \ra \; | \;\; x \in \real^n \},
\ee
where $f \in \real^n$ is a given vector, $\la x , f \ra$ represents
the inner product in $\real^n$,  and
 $W:\real^n\rightarrow \real$  is a fourth order polynomial of the form
\[
W(x):= \frac{\alp}{2}\left(\frac{1}{2}|x|^2-\lambda\right)^2 ,
\]
in which, $\alp$ and $\lambda$ are given positive parameters.\\

\begin{figure*}[h]
  \centering
  \subfloat[Function $\Pi$ when $n,f,\alp=1$ and
  $\lam=3$.]{\label{figsbs1}\includegraphics[width=0.4\textwidth]{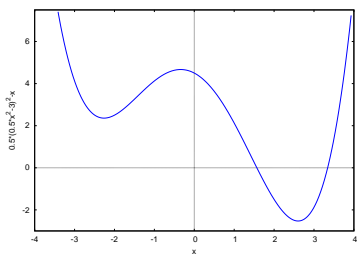}}\quad
  \subfloat[Function $\Pi$ when $n=2$, $f=0$, $\alp=1$ and $\lam=3$.]
  {\label{figsbs2}\includegraphics[width=0.4\textwidth]{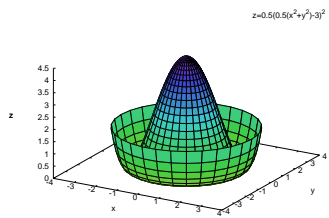}}
  \caption{Graphs of $\Pi(x)$} \label{figsbs}
\end{figure*}

 The non-convex problem $(\calP)$  appears extensively in many applications of
 sciences and engineering. For example, in the case that
$n=1$, $\Pi(x)$ is a {\em double-well function} (see figure
\ref{figsbs1}), which was first studied by van der Waals in thermal
mechanics in 1895. If $n=2$ and $f=0$ this is the so-called Mexican
hat function (see Figure \ref{figsbs2}) in cosmology and
theoretical physics.
Due to the nonconvexity, the function  $\Pi(x)$ may
possess multiple critical points, determined by the
necessary condition
\[
\nabla \Pi(x) =  \alp \left( \half | x |^2 - \lam \right) x - f  = 0.
\]
Direct methods for solving this nonlinear algebraic equation are difficult,
and to identify the global minimizer is a main task in global optimization.

If instead of the function $W$ considered above, we were to consider the function $W_B:\real^n\rightarrow\real$ defined by $$W_B(x):=\frac{\alp}{2}\left(\frac{1}{2}|Bx|^2-\lambda\right)^2,$$
where $B:\real^n\rightarrow \real^m$ is a linear transformation (not identically zero), now the function $\Pi(x)=W_B(x)-\la f,x\ra$ gives a more general case for problem $(P)$. Yet, if we take $f\in
R(B^tB)$, we can always reduce $(P)$ to the case where $B=I$ in the following way: make $y=Bx$ and let $\overline{f}=B(B^tB)^\dagger f$, where $(B^tB)^\dagger$ is the Moore-Penrose pseudoinverse of $B^tB$ (see \cite{DESOER}, \cite{PETERS} and references therein). Notice that since $f\in R(B^tB)$ then $B^t\overline{f}=f$ and $$\la f,x\ra=\la B^t\overline{f},x\ra=\la \overline{f},Bx\ra =\la
\overline{f},y\ra.$$ With this, we can define the function $\Pi_B:R(B)\rightarrow \real$ where $$\Pi_B(y)=\frac{\alp}{2}\left(\frac{1}{2}|y|^2-\lambda\right)^2 -\la \overline{f},y\ra.$$

If $y_0\in R(B)$ is a solution of $\Pi_B$, there must exist a $x_0\in \real^n$ such that $Bx_0=y_0$, then $$B^tBx_0=B^ty_0$$ and from this, we can take $x_0=(B^tB)^\dagger B^ty_0$ as a solution for
$\Pi$. Thanks to this, we will study only the case when $B=I$.

Canonical duality theory developed in
 \cite{gao-book} is potentially powerful for solving a large class of
 nonconvex/nonsmooth/discrete   problems in both analysis and global optimization
 \cite{gao-cc09,gao-ogden-qjmam08,gao-ruan-jogo10}.
 This theory is composed mainly of (a) a canonical dual transformation,
 (b) a complementary-dual principle, and (c) a triality theory.
  It was shown in  \cite{gao-opt03} that by the  canonical dual transformation, the
fourth-order nonconvex problem $(\calP)$ is equivalent to an
one-dimensional canonical dual problem which can be solved
analytically to obtain all critical points.
The complementary-dual principle shows that a complete set of  solutions
to  the primal problem can be represented analytically by these canonical dual solutions.
By the triality theory, both global minimizer and local maximizer can
be identified.
However, it was discovered in 2003 that
in order to identify local minimizer, the triality theory proposed in
 \cite{gao-book}
needs ``certain additional constraints" (see Remark 1 in \cite{gao-opt03}).
Therefore, the double-min duality statement in this triality theory was left as an open problem
in global optimization \cite{gao-amma03}.

The canonical duality theory and the associated triality have been challenged recently by
 Voisei and  Z$\breve{a}$linescua  in a set of
more than seven papers\footnote{See the web page at
http://www.math.uaic.ro/~zalinesc/reports.htm }.
Unfortunately, in these papers, they either made mistakes in
understanding some basic terminologies of  finite deformation mechanics, or
repeatedly  address  the same type of open problem for the double-min duality
 left unaddressed by Gao in
\cite{gao-opt03,gao-amma03}.
For example, the  {\em external energy} $F(u)$
  in conservative systems (the case studied by Gao and Strang in
  \cite{gao-gs1})  means that the gradient $\nabla F(u)$ must be a given external force field.
  Therefore, the function(al)  $F(u)$ in Gao and Strang's work can not be quadratic.
However,  in  their paper published recently in {\em Applicable Analysis},
quadratic $F(u)$ has been used by
 Voisei and  Z$\breve{a}$linescua in all   ``counterexamples".
Also, interested readers should find that the
references  \cite{gao-opt03,gao-amma03}, where the open problem was remarked,
 never been cited in any one of their papers.

The main purpose of this paper is to  solve  this open problem such that
the proposed  problem
  $(\calP)$ can be solved completely.
  The method and results presented in this paper have been used to prove the
  triality theory   for   global optimization problems with general polynomials \cite{gao-wu-jogo11}
  and general objective functions \cite{gao-wu-JOGO,silva-gao-JMAA2}.

\section{Canonical Dual Problem and Analytical Solutions}

Following the standard procedure of the canonical dual transformation,
first  we need to choose a geometric operator
$\Lam:\real^n\rightarrow [-\lam,+\infty)$ given by the following function
$$\Lam(x)=\frac{1}{2}|x|^2-\lam,$$
and the associated canonical function
$V:[-\lam,+\infty) \rightarrow \real_+$ defined by
$$V(\xi)=\frac{\alp}{2}\xi^2.$$
Therefore, the primal function  $\Pi$ can be reformulated as
$$\Pi(x)=V(\Lam(x))-\la f,x\ra.$$

By the Legendre transformation (see \cite{ARNOLD,SEWELL,ZIA}), the conjugate function
  $V^c:[-\alp\lam,+\infty)\rightarrow
\real_+$ is given by
$$V^c(\vsig)=\frac{\vsig^2}{2\alp}.$$
With this, the Gao-Strang {\em total complementary function}
$\Xi:\real^n\times[-\alp\lam,+\infty)\rightarrow \real$,
associated to the problem $(\calP)$ can be defined as follows:
$$\Xi(x,\vsig)=\vsig\Lam(x)-\frac{\vsig^2}{2\alp}-\la f,x\ra.$$
 Via this $\Xi(x, \vsig)$, the
canonical dual function
$\Pi^d:[-\alp\lam,+\infty)\rightarrow\real$ can be finally obtained by
\cite{gao-opt03}
\begin{eqnarray*}
\Pi^d(\vsig) & := & \staxn\Xi(x,\vsig)=\{\Xi(x_0 ,\vsig) | \;\;
\nabla_x\Xi(x_0 ,\vsig)=0\}\\
 & =  & -\frac{|f|^2}{2\vsig} -\frac{\vsig^2}{2\alp}-\vsig \lam,\ \forall\vsig\in
[-\alp\lam,+\infty),
\end{eqnarray*}
where the notation $\sta \{ * \}$ stands for finding stationary points of the function given in $\{ * \}$.

Notice that if $\vsig>0$ then the dual function $\Pi^d(\vsig)$
 is strictly concave which admits a unique
global maximizer;
however,   $\Pi^d (\vsig)$
is a d.c. function (difference of convex functions) on $[-\alp \lam, 0)$,
which should give us
  information about local extrema of the primal function $\Pi$.
  Therefore, the canonical dual problem is proposed in the following
  stationary form:
  \eb
  (\calP^d): \;\;\; {\sta}_{\hspace{-.82cm}{}_{{}_{{}_{\vsig\in[-\alp \lam, + \infty)}}}}\Pi^d(\vsig)=\{\Pi^d(\vsig_o)| \nabla\Pi^d(\vsig_o)=0\}.
  \ee

By the fact that the canonical dual problem $(\calP^d)$
has only one variable, the criticality condition
$\nabla \Pi^d(\vsig) = 0$, where
\begin{equation}\label{DoPId}
\nabla\Pi^d(\vsig)=\frac{|f|^2}{2\vsig^2}-\frac{\vsig}{\alp}-\lam,
\end{equation} leads to a quebec algebraic equation
\begin{equation}\label{sigmaroots}
2\vsig^2\left(\frac{\vsig}{\alp}+\lam\right)=|f|^2,
\end{equation}
which can be solved explicitly to obtain all three possible real solutions:
\begin{eqnarray}
\label{vsig1MAXim}\vsig_1 & = & r^{1/3}+\frac{\alp^2\cdot\lam^2}{9r^{1/3}}-\frac{\alp\cdot\lam}{3}\\
\vsig_2 & = & \left( -\frac{\sqrt{3}\,i}{2}-\frac{1}{2}\right)\cdot
r^{1/3}+\frac{\left(\frac{\sqrt{3}\,i}{2}-\frac{1}{2}\right)\cdot\alp^2\cdot\lam^2}{9r^{1/3}}
-\frac{\alp\cdot\lam}{3} \\
\label{vsig3MAXim}\vsig_3 & = &
\left(\frac{\sqrt{3}\,i}{2}-\frac{1}{2}\right)\cdot
r^{1/3}+\frac{\left(
 -\frac{\sqrt{3}\,i}{2}-\frac{1}{2}\right)\cdot\alp^2\cdot\lam^2}{9r^{1/3}}
-\frac{\alp\cdot\lam}{3},
\end{eqnarray}
where
\[
r=\frac{\alp\cdot|f|\,\sqrt{27\,{|f|}^{2}-8\,{\alp}^{2}\cdot{\lam}^{3}}}{4\cdot{3}^{\frac{3}{2}}}
+\frac{27\,{\alp}\cdot{|f|}^{2}-4\,{\alp}^3\cdot{\lam}^{3}}{108}.
\]
It is not difficult to show that if $|f|^2>8\alp^2\lam^3/27$ then $\vsig_1$ is the only real positive root and if $|f|^2=8\alp^2\lam^3/27$ then
$\vsig_1$ is positive, and $\vsig_2=\vsig_3=-2\alp\lam/3$. If $0<|f|^2<8\alp^2\lam^3/27$, equations \eqref{vsig1MAXim}-\eqref{vsig3MAXim} can be simplified further to obtain:
\begin{eqnarray}
\label{vsig1MAX}\vsig_1 & = & \frac{\alp\lam}{3}\left(2\cos\left(\frac{1}{3}\cos^{-1}\left(\frac{27|f|^2}{4\alp^2\lam^3}-1\right)\right)-1\right)\\
\vsig_2 & = & \frac{\alp\lam}{3}\left(2\cos\left(\frac{1}{3}\cos^{-1}\left(\frac{27|f|^2}{4\alp^2\lam^3}-1\right)+\frac{4\pi}{3}\right)-1\right) \\
\label{vsig3MAX}\vsig_3 & = & \frac{\alp\lam}{3}\left(2\cos\left(\frac{1}{3}\cos^{-1}\left(\frac{27|f|^2}{4\alp^2\lam^3}-1\right)+\frac{2\pi}{3}\right)-1\right).
\end{eqnarray}
Moreover, we will have that $\vsig_1>0>\vsig_2>-2\alp\lam/3>\vsig_3>-\alp\lam$.

\begin{figure}[h]
\centering
\includegraphics[scale=.4]{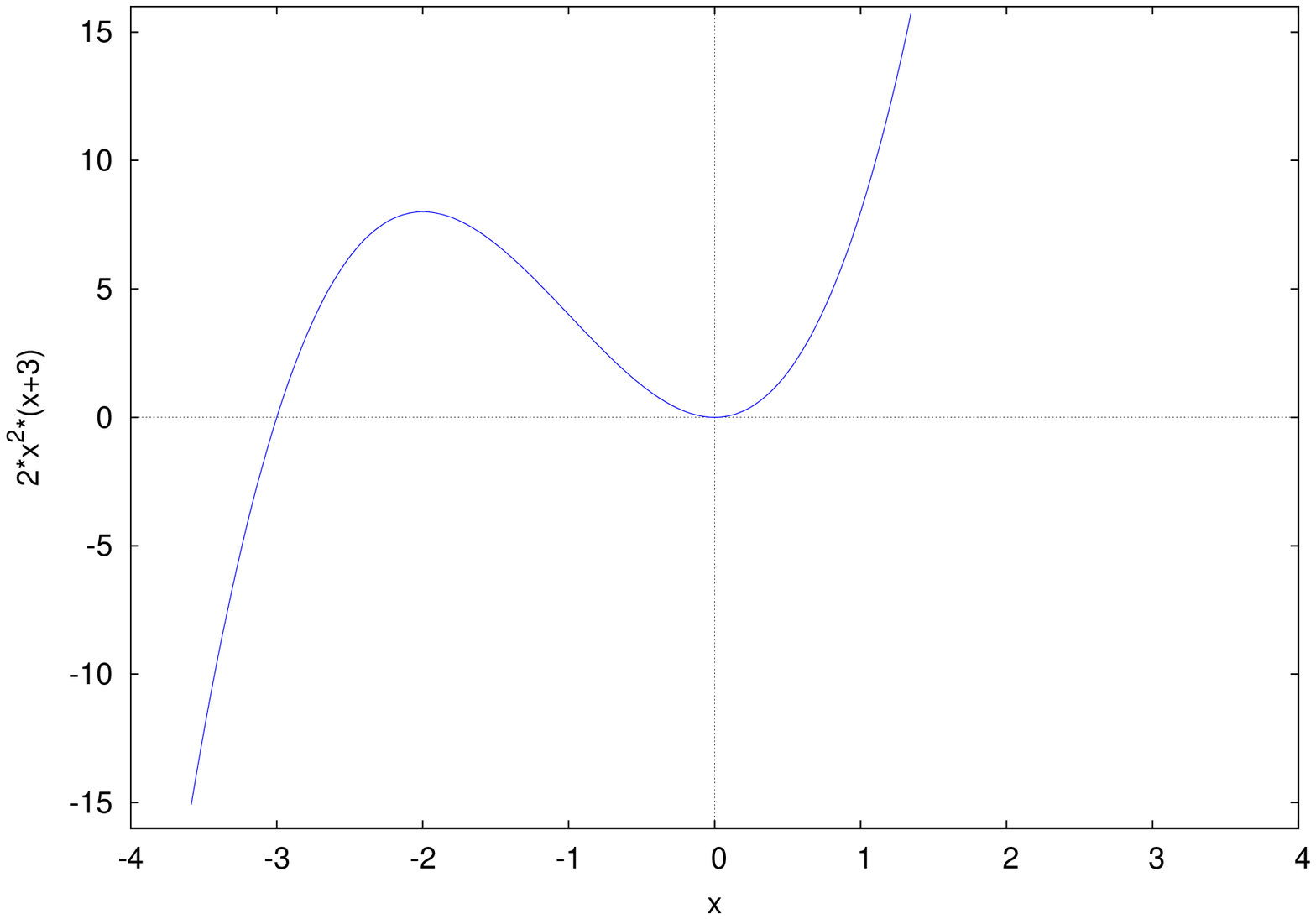}
\caption{The function $2\vsig^2(\vsig+\alp\lam)$ with $\alp\lam=3$.} \label{ggeq}
\end{figure}

\begin{thm}[Analytical Solutions \cite{gao-opt03}]
 Let $f\in \real^n\setminus\{0\}$ and $\{ \vsig_i\} $ be the real roots  of Equation
\eqref{sigmaroots}. Then, $x_i=f/\vsig_i$ are stationary points of
$(P)$ for every $i$ and $\Pi(x_i)=\Pi^d(\vsig_i)$.
\end{thm}
\textbf{Proof:} Notice that
$\Lam(x_i)=\vsig_i/\alp$, we have
$$\Lam(x_i)=\frac{1}{2}|x_i|^2-\lam=\frac{|f|^2}{2\vsig_i^2}-\lam$$
but since $\vsig_i$ is solution of \eqref{sigmaroots}, we have
\begin{equation}\label{Lamieqvsigialp}
\Lam(x_i)=\left(\frac{\vsig_i}{\alp}+\lam\right)-\lam=\frac{\vsig_i}{\alp}.
\end{equation}
Now, if we differentiate the function $\Pi$ we will have \begin{equation}\label{DoPi}
\nabla\Pi(x)=\alp\Lam(x)x-f\end{equation} and so
$$\nabla\Pi(x_i)=\alp\Lam(x_i)x_i-f=\alp\cdot\frac{\vsig_i}{\alp}\cdot\frac{f}{\vsig_i}-f=0.$$ On the
other hand $$\Pi(x_i)=\frac{\alp}{2}\cdot\frac{\vsig_i^2}{\alp^2}-\frac{|f|^2}{\vsig_i}=
\frac{\vsig_i^2}{2\alp}-\left(\frac{2\vsig_i^2}{\alp}
+2\vsig_i\lam\right)=-\frac{3\vsig_i^2}{2\alp}-2\vsig_i\lam$$
and
$$\Pi^d(\vsig_i)=-\frac{|f|^2}{2\vsig_i}-\frac{\vsig_i^2}{2\alp}-\vsig_i\lam=
-\left(\frac{\vsig_i^2}{\alp}+\vsig_i\lam\right)-\frac{\vsig_i^2}{2\alp}-\vsig_i\lam
=-\frac{3\vsig_i^2}{2\alp}-2\vsig_i\lam , $$
so we have  $\Pi(x_i)=\Pi^d(\vsig_i)$ as expected. \hfill $\blacksquare$\\

Theorem 1 shows that the stationary points of the dual problem induce naturally stationary points of the primal with zero duality gap. Using \eqref{DoPi}, it can be seen that if any stationary point of $\Pi$ exists, it must be in the same direction of $f$. Therefore, by analyzing the function $W(rf)$ with $r\in \real$ it can be seen that the possible stationary points of $W(rf)$ satisfy the following equation:
\eb\label{eqrforf}
\alp r \left(\frac{r^2}{2}|f|^2-\lam\right)=1.
\ee
Since $f\neq 0$, then $r\neq 0$ and by substituting $r=1/\vsig$ in \eqref{eqrforf} we will have \eqref{sigmaroots}. Thus, problem $(\calP)$ has at most three critical points. In the next section, we will show that the extremality of these solutions can be identified by a refined triality theory.

\section{Triality Theory and Perturbation}
The following spaces are important for understanding the triality theory:
\begin{eqnarray}
\calX_\sharp &:=& \left\{ x \in\real^n | \; \;
\left|\frac{\la f, x \ra}{|f||x|}\right| <\sqrt{-\frac{\vsig_2}{2\vsig_2+2\alp\lam}}
\right\},\\
\calX_\flat &:= & \left\{ x \in\real^n | \;\;
\left|\frac{\la
f,x \ra}{|f||x|}\right| > \sqrt{-\frac{\vsig_2}{2\vsig_2+2\alp\lam}} \right\}.
\end{eqnarray}

\begin{thm}[Refined Triality Theory]\label{ThExtrC}
Let $f\in\real^n$ be a given vector such that $0<|f|^2<8\alp^2\lam^3/27$, $\{\vsig_i \}$ with $i=1,2,3$ the three real roots of Equation \eqref{sigmaroots} such that $\vsig_1>0>\vsig_2>-2\alp\lam/3>\vsig_3>-\alp\lam$ and let $x_i=f/\vsig_i$.
Then we have
\begin{enumerate}
\item[i)] $x_1$ is a global minimizer of $\Pi$,  $\vsig_1$ is a maximizer  of $\Pi^d$ in $(0,+\infty)$, and
    \eb
    \Pi(x_1) = \min_{x\in \real^n} \Pi(x) = \max_{\vsig \in (0, +\infty)} \Pi^d(\vsig) = \Pi^d(\vsig_1).
    \ee
\item[ii)] There exist $\calX_o$ and $I_o$ neighborhoods of $x_3$ and $\vsig_3$ respectively such that $x_3$ is a local maximizer of $\Pi(x)$ in $\calX_o$ and $\vsig_3$ is a local maximizer of $\Pi^d(\vsig)$ in $I_o$, and
\eb
 \Pi(x_3) = \max_{x\in \calX_o} \Pi(x) = \max_{\vsig \in I_o} \Pi^d(\vsig) = \Pi^d(\vsig_3).
    \ee

\item[iii)] There exists $I_1$ a neighborhood of $\vsig_2$ such that $\vsig_2$ is a local minimizer of $\Pi^d(\vsig)$ in $I_1$ and $x_2$ is a saddle point of $\Pi(x)$. Specifically, $x_2$ is a local maximizer of $\Pi(x)$ in the
directions $x_2+ t \calX_\sharp$ and a local minimizer of $\Pi(x)$ in the directions  $x_2+ t \calX_\flat$, i.e.,
\eb
  \Pi(x_2) = \max_{t \in \real } \Pi(x_2 + t \calX_\sharp) = \min_{\vsig \in I_1} \Pi^d(\vsig) = \Pi^d(\vsig_2),
  \ee
  \eb
  \Pi(x_2) = \min_{t \in \real } \Pi(x_2 + t \calX_\flat) = \min_{\vsig \in I_1} \Pi^d(\vsig) = \Pi^d(\vsig_2).
  \ee
\end{enumerate}
\end{thm}
\textbf{Proof:}
\begin{enumerate}
\item[i)] The canonical dual solution $\vsig_1$ is a global minimizer  of $\Pi^d$ in $(0,+\infty)$
since $\Pi^d$ is a strictly concave function and $\vsig_1$ is its
only critical point in $(0,+\infty)$. Since $\vsig_1>0$,
$\Xi(\cdot,\vsig_1)$ is a strictly convex function, then its only
minimizer happens at its stationary point which is $x_1$. Also,
$\Xi(x,\vsig_1)\leq \Pi(x)$ for every $x$; in fact, since $V$ is a strictly convex function, by Fenchel's inequality for Convex functions we have that for every $\vsig$ and every $\xi$ $$\vsig\cdot\xi\leq V(\xi)+V^c(\vsig).$$ Taking $\vsig=\vsig_1$ and $\xi=\Lam(x)$ $$\vsig_1\Lam(x)\leq V(\Lam(x))+V^c(\vsig_1),$$ rearranging the last inequality and adding $-\la f,x\ra$ to both sides we have $\Xi(x,\vsig_1)\leq \Pi(x)$ for every $x\in \real^n$.\\

%take an arbitrary $x\in \real^n$:
%$$0\leq\frac{\alp}{2}\left(\Lam(x)-\frac{\vsig_1}{\alp}\right)^2=\frac{\alp}{2}(\Lam(x))^2-\vsig_1\Lam(x)+\frac{\vsig_1^2}{2\alp}$$
%$$\Longrightarrow \vsig_1\Lam(x)-\frac{\vsig_1^2}{2\alp}\leq \frac{\alp}{2}(\Lam(x))^2$$
%by adding $-\la f,x\ra$ to both sides of the last inequality, we have $\Xi(x,\vsig_1)\leq \Pi(x)$.

Using Equation \eqref{Lamieqvsigialp}, $\Lam(x_1)=\vsig_1/\alp$, it can be easily shown that
$\Pi(x_1)=\Xi(x_1,\vsig_1)$. With this, assume that there exists
$x'\in\real^n$ such that $\Pi(x_1)>\Pi(x')$ then
$$\Pi(x_1)>\Pi(x')\geq \Xi(x',\vsig_1)\geq \Xi(x_1,\vsig_1)=\Pi(x_1)$$
which is a contradiction. Therefore $x_1$ is a solution of $(P)$.

\item[ii)] By  the second derivative of $\Pi^d$,  we have:
$$\nabla^2\Pi^d(\vsig)=-\frac{|f|^2}{\vsig^3}-\frac{1}{\alp}.$$
 Then
\begin{eqnarray}
\nabla^2 \Pi^d(\vsig_i) & = & -\frac{|f|^2}{\vsig_i^3} -\frac{1}{\alp}=-\frac{2}{\vsig_i}\left(\frac{\vsig_i}{\alp}
+\lam\right)-\frac{1}{\alp} \nonumber \\
& = & -\frac{2}{\alp}-\frac{2\lam}{\vsig_i}-\frac{1}{\alp}
=\frac{3\vsig_i+2\alp\lam}{-\alp\vsig_i}.\label{D2Pd}
\end{eqnarray}

For $i=3$, $\nabla^2\Pi^d(\vsig_3)<0$ and $\vsig_3$ is a
local maximizer of $\Pi^d$.\\

On the other hand, by differentiating \eqref{DoPi} we have:
$$\nabla^2\Pi(x)=\alp\left(xx^t+\Lam(x)I\right)$$ and
$$\nabla^2\Pi(x_i)=\alp\left(x_ix^t_i+\Lam(x_i)I\right)=\alp\left(\frac{ff^t}{\vsig_i^2}+\frac{\vsig_i}{\alp}I\right).$$
For $i=3$, take any $z\in \real^n$:
$$z^t\nabla^2\Pi(x_3)z=\alp\left(\frac{z^tff^tz}{\vsig_3^2}+\frac{\vsig_3}{\alp}|z|^2\right)=\alp\left(\frac{\la
f,z\ra^2}{\vsig_3^2}+\frac{\vsig_3}{\alp}|z|^2\right),$$
therefore, by
  the Cauchy-Schwarz
inequality we have
\begin{eqnarray}
z^t\nabla^2\Pi(x_3)z  & \leq &
\alp\left(\frac{|f|^2|z|^2}{\vsig_3^2}+\frac{\vsig_3}{\alp}|z|^2\right)
=\alp\cdot|z|^2\left(
\frac{|f|^2}{\vsig_3^2} +\frac{\vsig_3}{\alp} \right) \nonumber \\
&=& \alp\cdot|z|^2\left(\frac{3\vsig_3}{\alp}+2\lam\right).\label{ztD2Pz}
\end{eqnarray}
But the expression in brackets is negative so $z^t\nabla^2\Pi(x_3)z\leq
0$ for every $z\in \real^n$ and $\Pi$ has a local maximizer at
$x_3$.

\item[iii)] Using Equation
\eqref{D2Pd} with $i=2$, we have that
$\nabla^2\Pi^d(\vsig_2)>0$ and $\vsig_2$ is a local minimizer
of $\Pi^d$.\\

On the other hand, by taking $z\in \real^n$, we know that $\phi(t)=\Pi(x_2+tz)$
has first and second derivatives as follows:
$$\phi'(t)=\nabla\Pi(x_2+tz)z,\quad\phi''(t)=z^t\nabla^2\Pi(x_2+tz)z.$$
Clearly, $\phi'(0)=0$. What about $\phi''(0)$?\\ \\ Consider
$\theta$ the angle between $z$ and $f$. Then
$$\phi''(0)=z^t\nabla^2\Pi(x_2)z=\alp\left(\frac{\la f,z\ra^2}{\vsig_2^2}+\frac{\vsig_2}{\alp}|z|^2\right)=
\alp\left(\frac{|f|^2|z|^2\cos^2\theta}{\vsig_2^2}+\frac{\vsig_2}{\alp}|z|^2\right)$$
$$=\alp|z|^2\left(\frac{|f|^2}{\vsig_2^2}\cos^2\theta+\frac{\vsig_2}{\alp}\right)=
\alp|z|^2\left(\left(\frac{2\vsig_2}{\alp}+2\lam\right)\cos^2\theta+\frac{\vsig_2}{\alp}\right),$$ so
\begin{equation}\label{D2Wsig2}
\phi''(0)=|z|^2\left((2\vsig_2+2\alp\lam)\cos^2\theta+\vsig_2\right).
\end{equation}

If $z\in \calX_\sharp$, by the definition of $\calX_\sharp$, we have that
$$|\cos\theta|= \left|\frac{\la f, x \ra}{|f||x|}\right| < \sqrt{-\frac{\vsig_2}{2\vsig_2+2\alp\lam}}.$$ Then $$\cos^2\theta<
-\frac{\vsig_2}{2\vsig_2+2\alp\lam}$$ $$ (2\vsig_2+2\alp\lam)\cos^2
\theta<-\vsig_2$$ \begin{equation}\label{sig2th1}(2\vsig_2+2\alp\lam)\cos^2\theta+\vsig_2<0.\end{equation}

So, substituting \eqref{sig2th1} into \eqref{D2Wsig2} implies that,
$\phi''(0)<0$ and $t=0$ is a local maximizer.\\

If $z\in \calX_\flat$, then by definition, we have
$$\sqrt{-\frac{\vsig_2}{2\vsig_2+2\alp\lam}}< \left|\frac{\la f, x \ra}{|f||x|}\right| = |\cos\theta|.$$ Then
$$ -\frac{\vsig_2}{2\vsig_2+2\alp\lam}< \cos^2\theta,$$ and this implies that
\begin{equation}\label{sig2th2}0<
(2\vsig_2+2\alp\lam)\cos^2\theta+\vsig_2.\end{equation}
 Thus,  from the equation  \eqref{D2Wsig2}
we know that $\phi''(0)>0$ and $t=0$ is a local minimizer.\hfill $\blacksquare$

\end{enumerate}
\textbf{Remark 1:} The triality theory says precisely that if
$\vsig_1$ is a global maximizer of $\Pi^d$ on a certain set, then
$x_1$ is a global minimizer for $\Pi$. This is known from the general result
by Gao and Strang in \cite{gao-gs1}.
If $\vsig_3$ is a
local maximizer for $\Pi^d$ then $x_3$ is also a local maximizer for
$\Pi$. This is the so-called double-max duality statement.
   If $\vsig_2$ is a local minimizer for $\Pi^d$, then
$x_2$ is also a local minimizer for $\Pi$ in certain directions.
This is so-called double-min duality in the standard triality form proposed in
\cite{gao-book}. The ``certain additional constraint"
discovered  in \cite{gao-opt03,gao-amma03} is
$ x = x_2+ t \calX_\flat \;\; \forall t \in \real$.
 Part iii of Theorem \ref{ThExtrC} is showing that
$x_2$ is, in fact, a saddle point. This solves the open problem left
in \cite{gao-opt03,gao-amma03} for this special case of double-well potential problem.  \\
%In other words, if
% $\Pi$ was to have a local minimum it should happen at $x_2$.\\ \\ %By perturbing the function $W$ in only one direction, namely $f$,
%and choosing a geometric operator $\Lam$ which has domain in
%$\real^n$ and range in $\real$, we can not control $n$ dimensions in
%an effective manner. Notice also that the function $W$ has infinite
%many global minimizers while $\Pi$ only has one.\\
\textbf{Remark 2:} If $|f|^2=8\alp^2\lam^3/27$, then
$\vsig_2=\vsig_3=-2\alp\lam/3$ and Equation \eqref{D2Pd} implies
that $\nabla^2\Pi^d(-2\alp\lam/3)=0$, even more, it is not hard to show
that this is an inflexion point of $\Pi^d$. The triality theory in
this case can not tell us what kind of stationary point is for $x_2$.
However, Equation \eqref{ztD2Pz} remains true, and in this case
(recall that $x_3 = x_2$) the expression in brackets is zero. So this
implies that $z^t\nabla^2\Pi(x_2)z\leq 0$ for every $z\in \real^n$ and
$x_2$ is a local maximizer of $\Pi$.\\

%By Equation \eqref{DoPi} we can see that the only critical points of
%$\Pi$ are in the direction of $f$, and thanks to Equation
%\eqref{sigmaroots}

It is clear that if  $f = 0$, the problem $(\calP)$ has infinite number of
global minimizers, they all lie in the sphere $|x|^2=2\lam$. In this case, the canonical dual is strictly concave with
only one local maximizer $\vsig_2$, which leads to a local maximizer $x=0$ of the primal problem.
 Therefore, a linear perturbation method has been
introduced in \cite{ruan-gao-jiao-coap} for solving some NP-hard problems in global optimization.
The next theorem proves the convergence of this canonical dual perturbation method under the current setting. Notice that we want to find at least a solution of $(\calP)$ if $f=0$.

\begin{thm}
Consider $(\calP)$ with $f=0$. Let $\alp,\lam\in \real^+$, $f_o\in\real^n$ such that $0<|f_o|^2<8\alp^2\lam^3/27$ and
consider $f_k=f_o/k,$ for every $k\in I\!\!N$. For $i=1,2,3$, take $\vsig_{i,k}$ the
critical points of $\Pi_k^d$ which is the dual function induced by $\Pi_k(x)=W(x)- \la x,f_k\ra$, and $x_{i,k}=f_k/\vsig_{i,k}$. Then
$$\lim_{k\rightarrow \infty}x_{1,k}=\overline{x}_1\text{ and }
|\overline{x}_1|^2=2\lam,$$
$$\lim_{k\rightarrow \infty}x_{2,k}=\overline{x}_2\text{ and }
|\overline{x}_2|^2=2\lam,$$
$$\lim_{k\rightarrow \infty}x_{3,k}=0.$$
\end{thm}
\textbf{Proof:} Since $x_{i,k}=f_k/\vsig_{i,k}=f_o/(k\vsig_{i,k})$ we need
to show that $1/(k\vsig_{i,k})$ converges for every $i$. Since $f_k$
is converging to zero, from equations \eqref{vsig1MAX}-\eqref{vsig3MAX}, we can see that $\vsig_{1,k}$ and $\vsig_{2,k}$ both converge to zero and $\vsig_{3,k}$ converges to $-\alp\lam$. Thanks to \eqref{sigmaroots} we know that
$$2\left(\frac{\vsig_{i,k}}{\alp}+\lam\right)=\frac{|f_k|^2}{(\vsig_{i,k})^2}= \frac{|f_o|^2}{(k\vsig_{i,k})^2},$$ which implies that $$\frac{1}{k|\vsig_{i,k}|}=\frac{\sqrt{2\left(\frac{\vsig_{i,k}}{\alp}+\lam\right)}}{|f_o|}.$$ With this, we have:

$$\lim_{k\rightarrow \infty}\frac{1}{k|\vsig_{1,k}|}= \lim_{k\rightarrow \infty}\frac{1}{k\vsig_{1,k}}=\lim_{k\rightarrow \infty} \frac{\sqrt{2\left(\frac{\vsig_{1,k}}{\alp}+\lam\right)}}{|f_o|}=\frac{\sqrt{2\lam}}{|f_o|},$$

$$\lim_{k\rightarrow \infty}\frac{1}{k|\vsig_{2,k}|}=- \lim_{k\rightarrow \infty}\frac{1}{k\vsig_{2,k}}=\lim_{k\rightarrow \infty} \frac{\sqrt{2\left(\frac{\vsig_{2,k}}{\alp}+\lam\right)}}{|f_o|}=\frac{\sqrt{2\lam}}{|f_o|},$$

and

$$\lim_{k\rightarrow \infty}\frac{1}{k|\vsig_{3,k}|}=- \lim_{k\rightarrow \infty}\frac{1}{k\vsig_{3,k}}=\lim_{k\rightarrow \infty} \frac{\sqrt{2\left(\frac{\vsig_{3,k}}{\alp}+\lam\right)}}{|f_o|}=\frac{0}{|f_o|}=0.$$ Finally, we
have $$\lim_{k\rightarrow \infty}x_{1,k}= \lim_{k\rightarrow \infty} \left(\frac{1}{k\vsig_{1,k}}\right)f_o=\sqrt{2\lam}\frac{f_o}{|f_o|},$$

$$\lim_{k\rightarrow \infty}x_{2,k}= \lim_{k\rightarrow \infty} \left(\frac{1}{k\vsig_{2,k}}\right)f_o=-\sqrt{2\lam}\frac{f_o}{|f_o|},$$ and

$$\lim_{k\rightarrow \infty}x_{3,k}= \lim_{k\rightarrow \infty} \left(\frac{1}{k\vsig_{3,k}}\right)f_o=0\cdot f_o=0.$$ With all this, we have just proven that $x_{1,k}$ and $x_{2,k}$ both converge to global minimizers of $W$, while $x_{3,k}$ converges to the local maximizer of $W$.\hfill $\blacksquare$

\section{Examples}

\subsection{Example 1: The case when $f=0$}

\begin{figure}
  \centering
  \subfloat[$\Pi(x,y)=0.5(0.5(x^2+y^2)-3)^2$.]{\label{gex0a}\includegraphics[width=0.4\textwidth]{gex0a.eps}}\quad
  \subfloat[$\Pi^d(\vsig)=-0.5\vsig^2-3\vsig$.]
  {\label{gex0b}\includegraphics[width=0.4\textwidth]{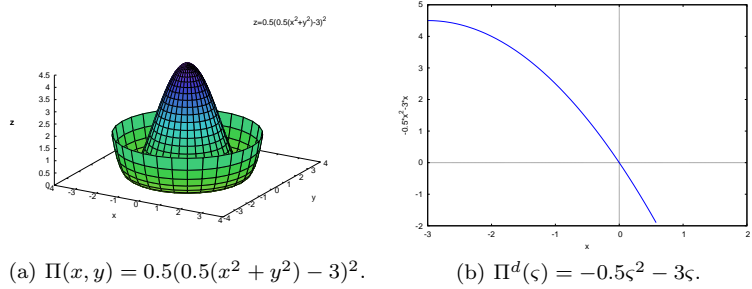}}
  \caption{Example 1} \label{gex0}
\end{figure}

Consider $n=2$, $\alp=1$ and $\lam=3$, just like in figure
\ref{figsbs2}. In this case, the dual function is given by
$\Pi^d(\vsig)=-0.5\vsig^2-3\vsig$. The graphs of the functions $\Pi$
and $\Pi^d$ are given in figure \ref{gex0}. Clearly, for $\Pi$, the
local maximizer is at the origin and the global minimizers are in
the sphere $|x|^2=6$. While $\Pi^d$ does not have stationary points.

\subsection{Example 2: The case when $0<|f|^2<8\alp^2\lam^3/27$}

Consider $n=2$, $\alp=1$, $\lam=3$ and $f=(1,1)$. In this case, the
functions $\Pi$ and $\Pi^d$ are given in figure \ref{gex1}.

Using Equations \eqref{vsig1MAX}-\eqref{vsig3MAX}, it is not hard to
show that the three stationary points of $\Pi^d$ are $\vsig_1=
2\cdot\cos40^\circ -1,\ \vsig_2= 2\cdot\cos80^\circ -1$ and
$\vsig_3= 2\cdot\cos160^\circ -1$.\\

\begin{figure}[h]
  \centering
  \subfloat[$\Pi(x,y)=0.5(0.5(x^2+y^2)-3)^2-x-y$.]{\label{gex1c}\includegraphics[width=0.4\textwidth]{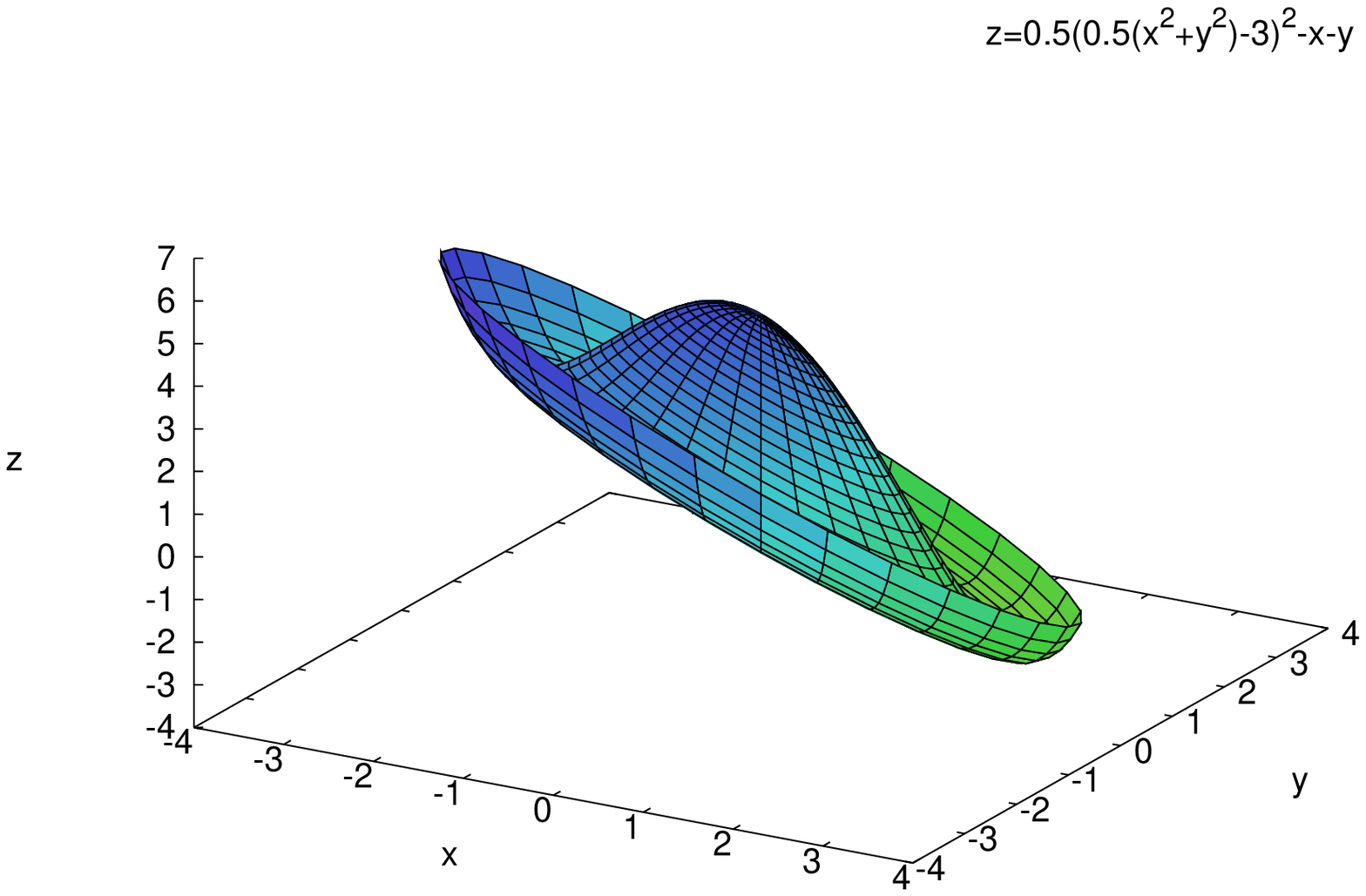}}\quad
  \subfloat[$\Pi^d(\vsig)=(-1/\vsig)-0.5\vsig^2-3\vsig$.]
  {\label{gex1d}\includegraphics[width=0.4\textwidth]{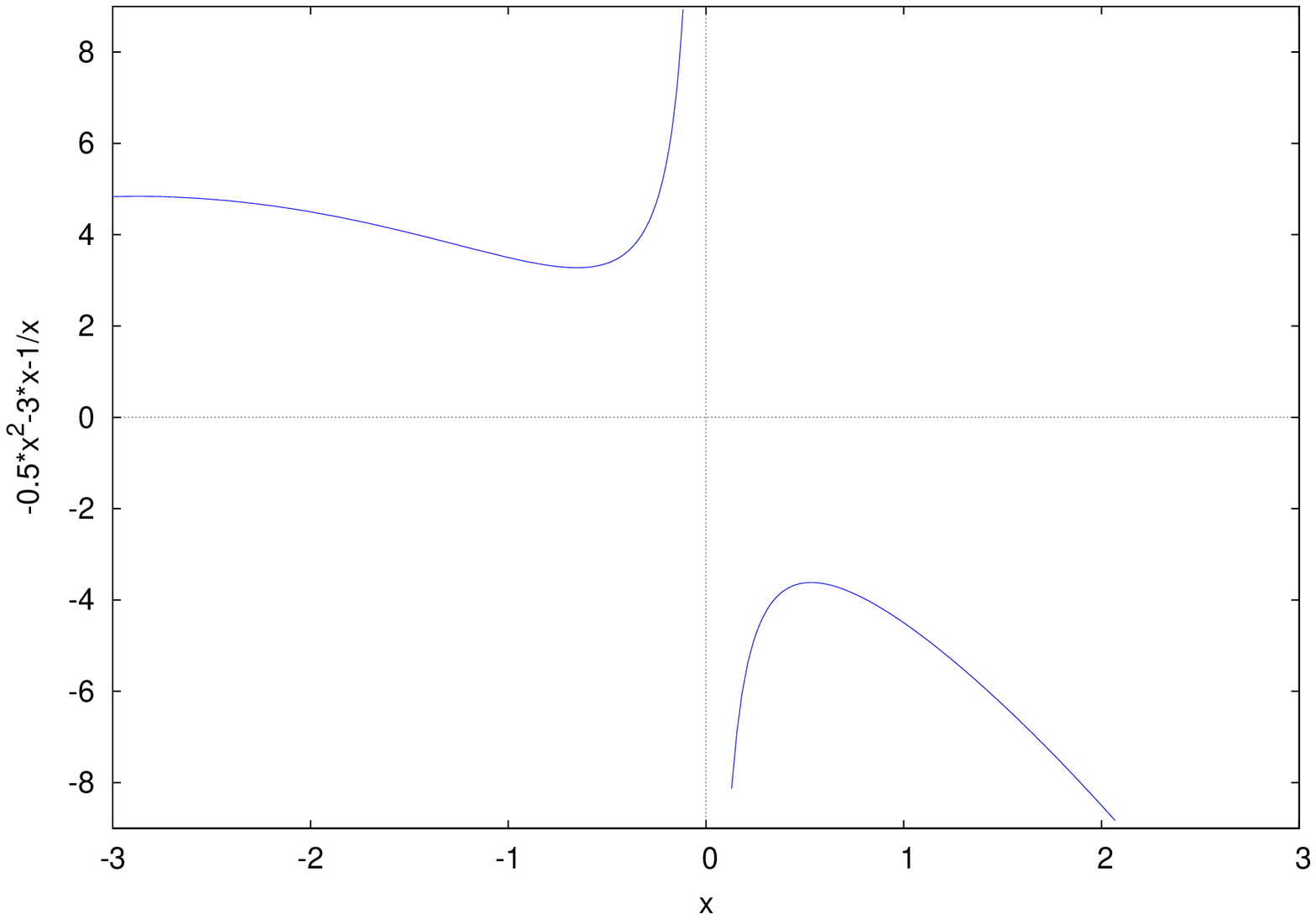}}
  \caption{Example 2} \label{gex1}
\end{figure}
\begin{figure}[h]
  \centering
  \subfloat[Function $\phi$ with $z=(1,-1)$.]{\label{gex1e}\includegraphics[width=0.4\textwidth]{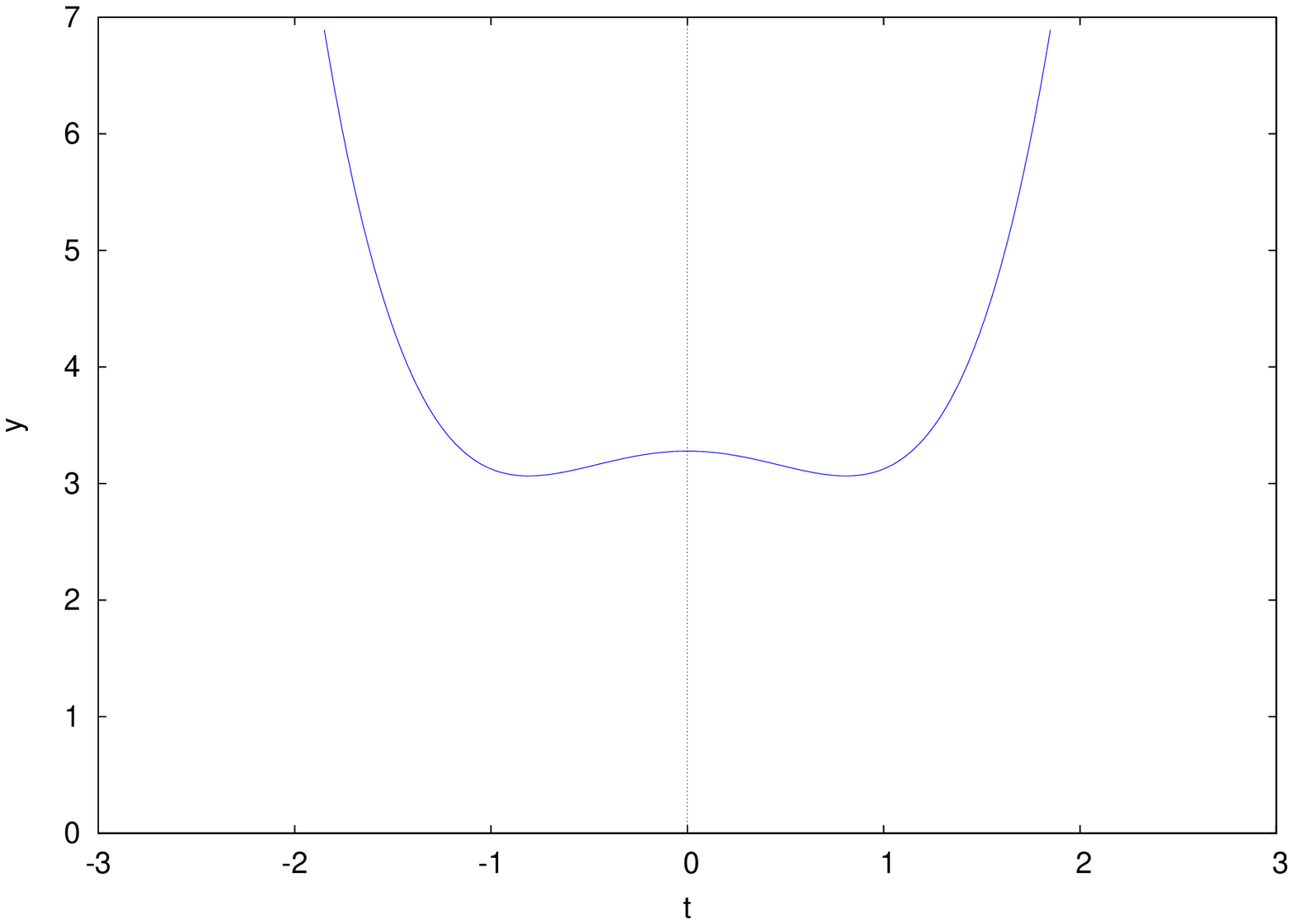}}\quad
  \subfloat[Function $\phi$ with $z=(0.2,1.4)$.]
  {\label{gex1f}\includegraphics[width=0.4\textwidth]{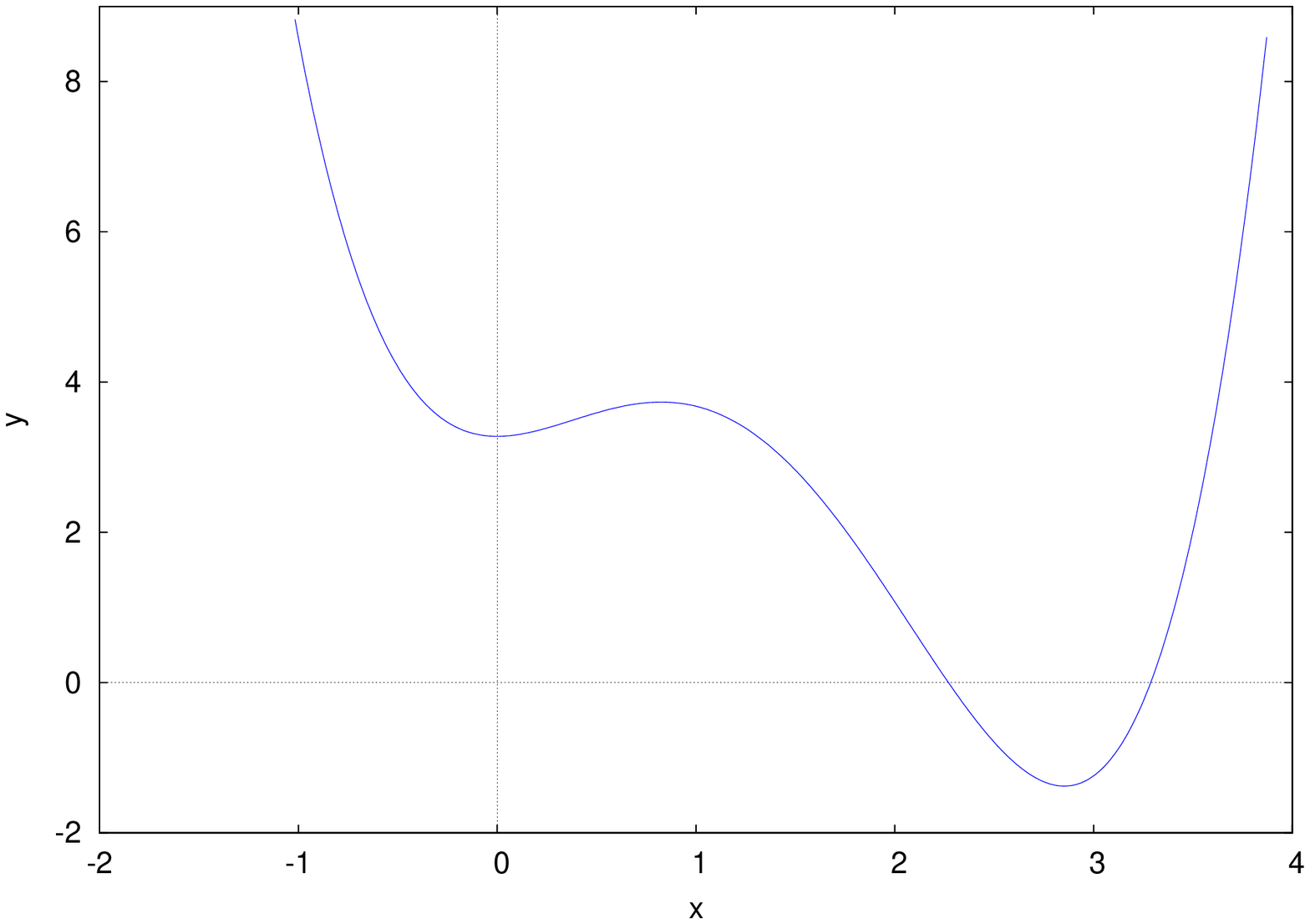}}
  \caption{Function $\phi$} \label{gex1}
\end{figure}

Let us highlight that in this case, $t=0$ could be a minimizer or a
maximizer of the function $\phi(t)=\Pi(x_2+tz)$, where
$x_2=f/\vsig_2$ and $z\in \real^n$ is an arbitrary chosen vector. If
we consider $z=(1,-1)$ we have that the graph of $\phi$ is given by
figure \ref{gex1e} and if we consider $z=(0.2,1.4)$ we have that the
graph of $\phi$ is given by figure \ref{gex1f}.

Clearly, $t=0$ is a local maximizer for $\phi$ if $z=(1,-1)$ and a
local minimizer if $z=(0.2,1.4)$.

\subsection{Example 3: The case when $|f|^2=8\alp^2\lam^3/27$}

\begin{figure}[h]
  \centering
  \subfloat[$\Pi(x,y)=0.5(0.5(x^2+y^2)-3)^2-2x-2y$.]{\label{gex3a}\includegraphics[width=0.4\textwidth]{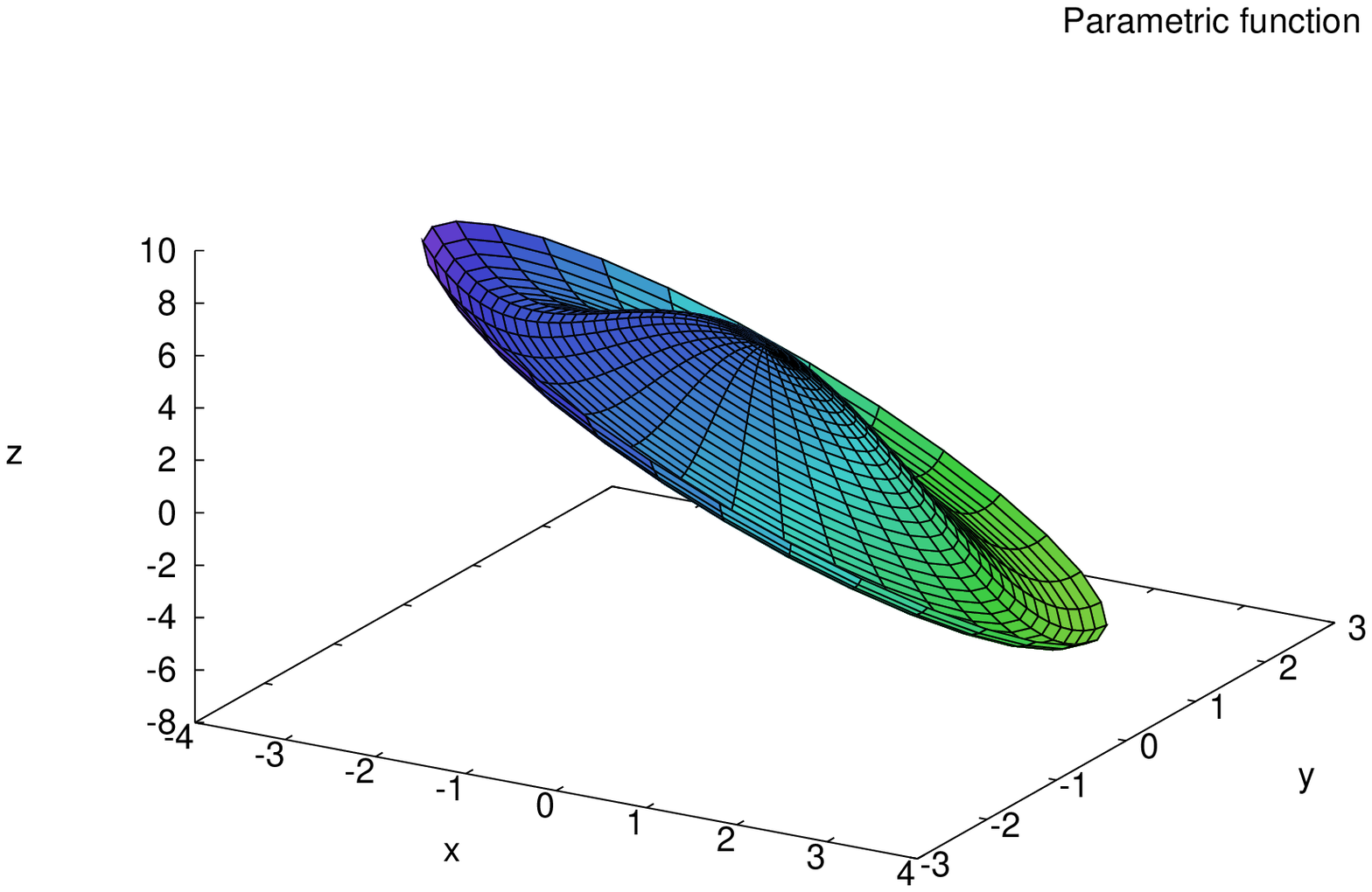}}\quad
  \subfloat[$\Pi^d(\vsig)=(-4/\vsig)-0.5\vsig^2-3\vsig$.]
  {\label{gex3b}\includegraphics[width=0.4\textwidth]{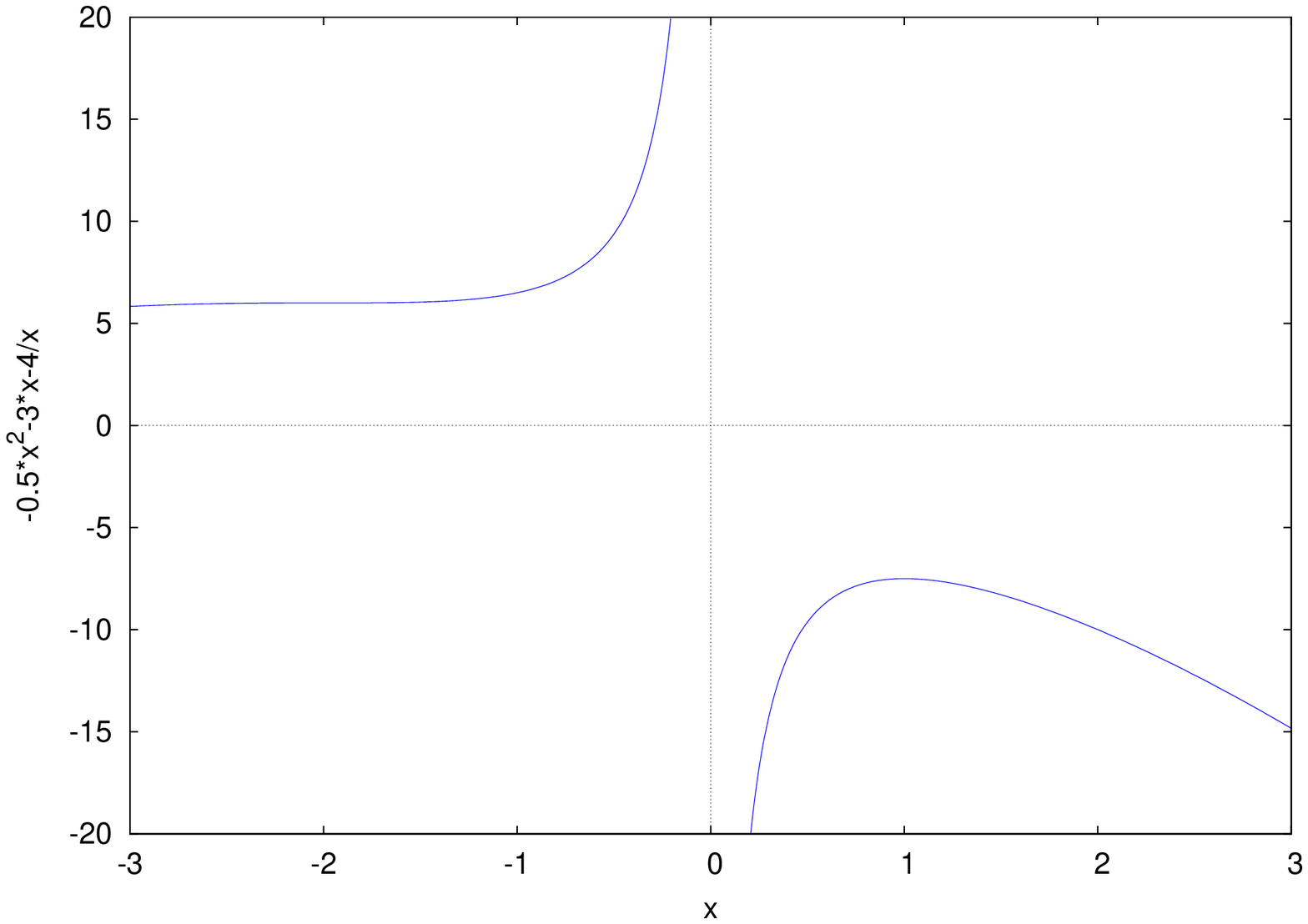}}
  \caption{Example 3} \label{gex3}
\end{figure}

Consider $n=2$, $\alp=1$, $\lam=3$ and $f=(2,2)$. In this case, the
functions $\Pi$ and $\Pi^d$ are given in figure \ref{gex3}.\\

Using Equations \eqref{vsig1MAX}-\eqref{vsig3MAX}, it is not hard to
show that the three stationary points of $\Pi^d$ are $\vsig_1=
\lam/3,\ \vsig_2=\vsig_3= -2\lam/3$.

\subsection{Example 4: The case when $8\alp^2\lam^3/27<|f|^2$}

\begin{figure}[h]
  \centering
  \subfloat[$\Pi(x,y)=0.5(0.5(x^2+y^2)-3)^2-3x-3y$.]{\label{gex4a}\includegraphics[width=0.4\textwidth]{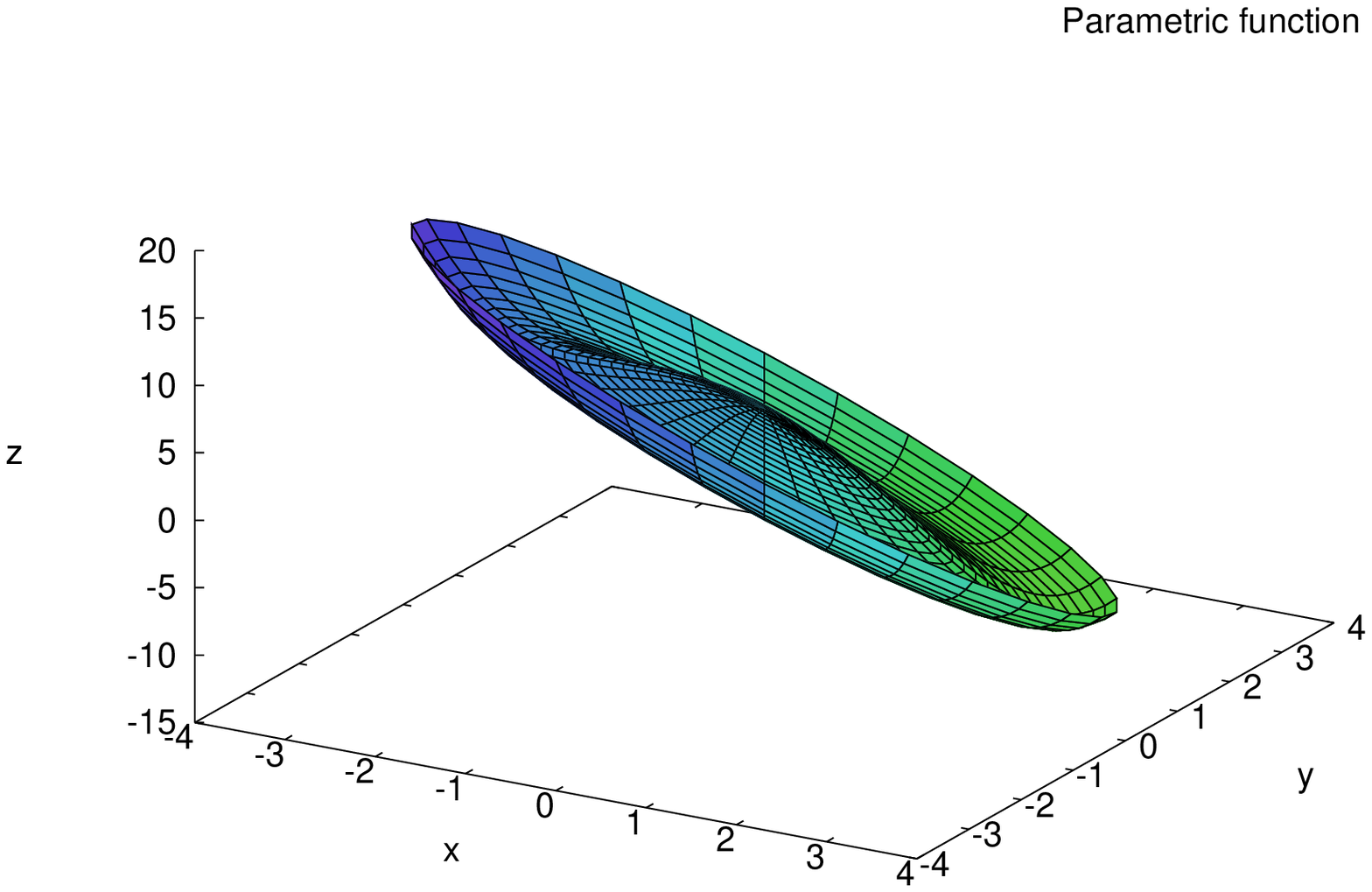}}\quad
  \subfloat[$\Pi^d(\vsig)=(-9/\vsig)-0.5\vsig^2-3\vsig$.]
  {\label{gex3b}\includegraphics[width=0.4\textwidth]{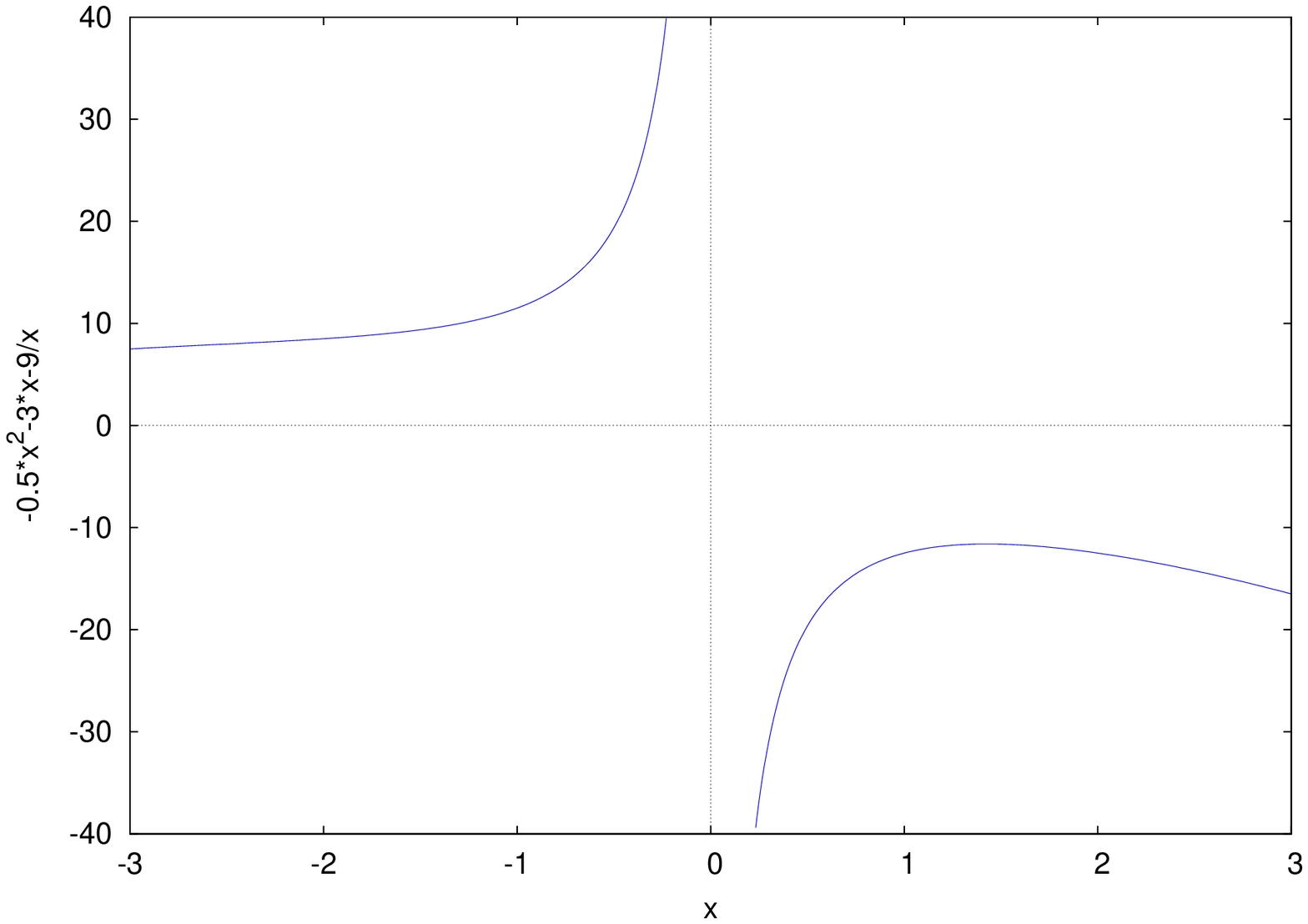}}
  \caption{Example 4} \label{gex4}
\end{figure}

Consider $n=2$, $\alp=1$, $\lam=3$ and $f=(3,3)$. In this case, the
functions $\Pi$ and $\Pi^d$ are given in figure \ref{gex4}.\\

From Equations \eqref{vsig1MAXim}-\eqref{vsig3MAXim}, it is not hard to
show that the only real stationary point of $\Pi^d$ is $\vsig_1$.\\

\section{Concluding  Remarks}
A complete set of analytical solutions is presented in this paper for a
nonconvex optimization problem with double-well potential in $\real^n$.
The open problem on the double-min duality left in 2003 has been solved for this special
case. But the method and idea developed in this paper
pave the way   to prove the triality theory
in general global optimization problems \cite{gao-wu-jogo11, gao-wu-JOGO, silva-gao-JMAA2}.
The perturbation  Theorem  3 shows that if the primal problem has more than one global minimizer,
the  linear  canonical dual perturbation method and the triality theory
can be used for finding  both global minimizer and local extrema.
It was first  realized in \cite{gao-jimo07} that the primal problem could be NP-hard if
it has more than one global minimizer.
Therefore, this linear perturbation method should play  a key role in solving
some challenging problems in global optimization  (see \cite{wang-fang-etal}).
Nonlinear perturbation method for solving NP-hard integer programming problems
has been discussed in \cite{gao-ruan-jogo10}.

\end{document}